\documentclass[11pt]{article}%
\usepackage{amssymb}
\usepackage{float}
\usepackage{amsfonts}
\usepackage{amsmath}
\usepackage[nohead]{geometry}
\usepackage[doublespacing]{setspace}
\usepackage[bottom]{footmisc}
\usepackage{indentfirst}
\usepackage{graphicx}%
\usepackage{rotating}
\usepackage[english]{babel}
\usepackage{lineno}
\usepackage{natbib}
\usepackage{textcomp}
\usepackage{array}
\usepackage{gensymb}
\usepackage{empheq}
\usepackage[anythingbreaks]{breakurl}
\usepackage{mathtools}
\usepackage{lscape}
\usepackage{systeme}

\usepackage{pdflscape}
\usepackage{dcolumn}

\usepackage{caption}
\usepackage{appendix}
\usepackage{subcaption}
\usepackage{hyperref}
\usepackage[T1]{fontenc}
\usepackage{booktabs} 
\usepackage{adjustbox,lipsum}
\usepackage{natbib}
\usepackage{footmisc}
\DefineFNsymbols{mySymbols}{{\ensuremath\dagger}{\ensuremath\ddagger}\S\P
   *{**}{\ensuremath{\dagger\dagger}}{\ensuremath{\ddagger\ddagger}}}
\setfnsymbol{mySymbols}
\setcitestyle{authoryear,open={},close={}}

\newcolumntype{L}{>{\centering\arraybackslash}m{3cm}}

\defcitealias{UN_SDG2016}{United Nations 2016}

\newcolumntype{C}[1]{>{\centering\arraybackslash}m{#1}}

\makeatletter
\def\@biblabel#1{\hspace*{-\labelsep}}
\makeatother
\begin{document}
\title{Anticipating Illegal Maritime Activities from Anomalous Multiscale Fleet Behaviors}
\author{James R. Watson and A. John Woodill\thanks{James R. Watson: College of Earth, Ocean and Atmospheric Sciences, Oregon State University; jrwatson@coas.oregonstate.edu. A. John Woodill: College of Earth, Ocean and Atmospheric Sciences, Oregon State University; johnwoodill@gmail.com. This work was supported by NASA under the award "Securing Sustainable Seas: Near real-time monitoring and predicting of global fishing fleet behavior" (Award No. 80NSSC19K0203).} \medskip\\ {\normalsize Oregon State University} }
\maketitle

\singlespacing
\begin{center}
\textbf{Abstract}
\end{center}

Illegal fishing is prevalent throughout the world and heavily impacts the health of our oceans, the sustainability and profitability of fisheries, and even acts to destabilize geopolitical relations. To achieve the United Nations Sustainable Development Goal of "Life Below Water", our ability to detect and predict illegal fishing must improve. Recent advances have been made through the use of vessel location data, however, most analyses to date focus on anomalous spatial behaviors of vessels one at a time. To improve predictions, we develop a method inspired by complex systems theory to monitor the anomalous multi-scale behavior of whole fleets as they respond to nearby illegal activities. Specifically, we analyze changes in the multiscale geospatial organization of fishing fleets operating on the Patagonia Shelf, an important fishing region with chronic exposure to illegal fishing. We show that legally operating (and visible) vessels respond anomalously to nearby illegal activities (by vessels that are difficult to detect). Indeed, precursor behaviors are identified, suggesting a path towards pre-empting illegal activities. This approach offers a promising step towards a global system for detecting, predicting and deterring illegal activities at sea in near real-time. Doing so will be a big step forward to achieving sustainable life underwater.

\strut


\strut

\vspace{1mm}
\begin{center}
\end{center}

\thispagestyle{empty}

\pagebreak%
\doublespacing

\section*{INTRODUCTION}

\noindent The United Nations Sustainable Development Goal (SDG) 14 is to conserve and sustainably use the ocean's resources for sustainable growth (\citepalias{UN_SDG2016}). This is a vital goal to achieve as many millions of people rely on the oceans for food and income, and unsustainable use of the seas will lead to diminished food and income security around the world (\cite{bene2016}). The SDG Target 14.4 is to have the global capacity to effectively regulate harvesting and end overfishing by 2020. In doing so, overharvested fish stocks will then be on a path towards recovery. Sustainable Development Goal 14 and its specific target can only be achieved if illegal, unreported and unregulated (IUU) fishing is dealt with (\cite{pramod2014}). This means being able to identify in (near) real-time IUU events around the world, and even predict their occurrences, all so that enforcement and intervention can occur more frequently and with more efficacy. Illegal fishing undermines SDG 14 by creating aggregate fishing effort above maximum sustainable levels and potentially creates geopolitical tension and possibly armed conflict, which have detrimental consequences for regional development (\cite{osterblom2012}). Since illegal fishers do not report their catch, their fishing activities reduce the accuracy of official fish catch and stock estimates and impedes the ability of regulatory bodies to set catch limits, manage fish populations, and evaluate SDG progress (\cite{worm2012}). Further, since their catch is not geographically referenced, their fishing activities undermine our global ability to monitor marine protection and the health of the oceans more broadly (\cite{lubchenco2015}). 

Equally important to IUU's socio-economic impacts are its effects on marine environments: IUU fishing often causes grave environmental damage, especially when vessels use prohibited gear, such as driftnets, that catch non-target species (like sharks, turtles or dolphins) or physically damages or destroys reefs, seamounts, and other vulnerable marine ecosystems. In concert, both the socio-economic and environmental impacts of IUU has a disproportionate effect on smaller-scale fishers in developing countries, by stealing fish from near-shore waters and undermining the ecosystem on which the fish depend (\cite{pauly2018}). Given that illegal fishing is valued at ~\$36.4 billion annually and represents 20\% of the global seafood catch (Sumaila et al. 2006; Stimson Center 2018), there is a pressing need to be able to detect illegal fishing in a timely manner and at spatial and temporal scales relevant to enforcement. In doing so, diminished IUU activity will lead to improved geopolitical relations in areas where IUU occurs, as well as increases in legal catch and profits (estimated around 14\% and 12\% respectively; \cite{cabral2018})

The biggest challenge to detecting and predicting IUU activity has to do with IUU vessels going dark, that is most "normal" vessels operating legally transmit their location to other vessels via the Automatic Information System (AIS; \cite{kroodsma2018}). This is done primarily for vessel collision avoidance and for search and rescue efforts. The trouble is that when certain vessels engage in illegal activities, they typically turn off their AIS transponders or even falsify their location (this is known as "spoofing"; \cite{desouza2016}). Looking across the globe there is a "global dark fleet" operating in and amongst the world's visible (legally operating) fleets (\cite{sumaila2006}) committing all sorts of bad behaviors, starting with illegal fishing but extending to narcotics, human and arms trafficking (\cite{tickler2018}). Most current approaches to detecting IUU fishing focuses on identifying spoofing in AIS data (e.g. \cite{ford2018}), which has led to several recent advances in our understanding of the global spatial distribution and possible scale of IUU fishing (\cite{miller2018}). However, these spoofing analyses are almost always retroactive, revealing anomalous behavior after they have occurred. To make predictions about where IUU activity might happen in the future, at spatial and temporal scales relevant to enforcement, new approaches are required. 

A major limit to current spoofing detection approaches is that they only analyze the anomalous spatial behavior of vessels one at a time (e.g. \cite{ford2018, miller2018, petrossian2018}). In contrast, studies of anomalous behaviors in complex systems show that early detection (and prediction) can be achieved through the analysis of multiscale patterns, that is a spatial anomaly (for example) that starts with one actor, but then that spreads to others (\cite{motter2004}). Identifying contagious behaviors has aided the discovery of large and abrupt changes in financial markets (\cite{gai2010}), ecosystems (\cite{levin1998}) and our bodies (e.g. the onset of an epileptic fit say; \cite{litt2002}). In the context of IUU fishing, the key conceptual advance is to acknowledge that any one vessel is embedded in a complex spatial system comprised of all the other vessels in its proximity. The expectation is that the spatial behavior of a given vessel is determined by their major objectives (e.g. moving from A to B, finding and catching fishing, etc.), but also by the proximity and behavior of nearby vessels. If a nearby vessel goes dark and commits an illegal act, then those nearby legally operating (and visible) vessels may respond anomalously, for instance, they may move in such a way as to avoid any trouble. 

For early detection and prediction of IUU fishing, this means not just looking for anomalous behaviors of any one vessel (i.e. our current abilities to detect spoofing), but in analyzing the anomalous behavior of whole fleets. A useful analogy is to visible matter in the universe, whose movement through space is in large part determined by unobserved dark matter. On the seas, the movement of visible fleets is determined in part by the actions of unobserved dark fleets. Thus, an approach based on the analysis of anomalous multiscale spatial patterns of visible fleets, as they react to nearby but unobserved vessels committing illegal activities, could result in an entirely new way to detect and even predict IUU fishing. Our goal here was to develop such an approach, analyzing AIS data using new multiscale anomaly detection algorithms inspired by similar approaches applied to complex systems. We applied our algorithms to vessel location data for the Patagonia shelf, a highly productive fisheries region, and reveal anomalous precursor signals for several IUU events in the area. These results offer a promising new path to pre-empting illegal behaviors at sea, with the ultimate goal of diminishing IUU activity worldwide, and achieving the Sustainable Development Goal 14: Life Under Water.

\section*{METHODS}

\noindent In order to develop new algorithms for detecting and predicting IUU fishing, we synthesized AIS data for a specific region of the world -- the Patagonia Shelf (see Figure 1 for a map of the region). The Patagonia shelf is one of the world's most productive areas for fisheries, and fleets from all over the world come to harvest many species (\cite{bisbal1995}).  Major fisheries include the illex squid, southern blue whiting and blue grenadier / hoki fisheries, and fleets that dominate the region include those from Argentina, the UK (through the Falklands Islands), China, and Spain. Indeed, in the last decade or so, the presence of Chinese fishing fleets has increased substantially (in line with their Ocean Silk Road initiative; \cite{sutter2012}). Along with increased fishing in general, the Patagonia Shelf has also experienced increased incidences of illegal fishing, in particular by Chinese fishing vessels. There have been a number of high-profile incidences where Chinese fishing vessels have been found to be fishing illegally, for example by the Argentine navy, and then either chased and even destroyed. To develop new abilities to detect and predict IUU activities, the Patagonia shelf and these few high profile IUU events serve as important case-studies with which to develop the new methodology.

\subsection*{Fishing Vessel Location Data}

\noindent Global AIS data for the period 2016-2018 were obtained through Global Fishing Watch\footnote{https://globalfishingwatch.org/datasets-and-code/}. AIS data were originally designed for collision avoidance at sea; vessels equipped with an AIS transponder transmit their position and vessel identification data, such as maritime mobile security information number (MMSI), call sign, ship type, speed and course over ground, and other information to ships nearby carrying similar transponders as well as to receiving ground stations and low-orbit satellites. Signal transmission frequencies vary with speeds between a few seconds and a few minutes. These high-resolution tracking data are synthesized by ORBCOMM, which were then obtained through Global Fishing Watch. For this analysis of IUU fishing, we subsetted these global data spatially to the Patagonia Shelf (Fig. 1) and temporally to specific historic IUU events in this area (see Table 1 for a description of these events).  

The AIS data required significant "cleaning" for it to be useable. Specifically, a significant fraction (12-20\%) of vessels were found to spoof their locations. That is specific vessel locations were observed to "jump" unrealistic distances, indicating anomalous AIS reporting. Another observed issue was multiple (non)unique identifiers (MMSI) reporting across the region (i.e. the impossible case where the same vessel is in two locations at the same time). To remove these erroneous data points, vessels with speeds greater than 32 kilometers per hour were removed. This threshold was calculated as the 99th percentile of the vessel speed distribution calculated over the whole region for the whole year of 2016. Vessels reporting locations on land, at a port or zero kilometers from shore were also removed from the data, as well as vessels that did not travel more than one kilometer per day. The remaining data were then interpolated to an hourly time-step to remove the intermittent reporting times by vessels. Although these cleaning steps removed a sizable fraction of the original data, for each IUU case-study on the Patagonia Shelf, there remained a usable number of vessel locations (see Table 1, last column).

\subsection*{Multiscale Fleet Spatial Behavior}

\noindent To quantify changes in the spatial behavior of entire fleets, we first calculated the "relationship" between vessels as simply their distances from one another. More specifically, for a given time (i.e. an hourly time-point in the processed data), for each vessel, the Haversine distance was calculated to all other vessels in the area (Fig. 1A; for clarity lines identify the nearest three vessels). Overall vessels and for a given time, this results in an NxN Haversine distance matrix, where N is the number of vessels in the region. From this distance matrix, a probability density function (PDF) is computed using kernel density estimation (e.g. see Fig. 1B). Overall time periods, this leads to a set of between vessel distance distributions, each summarizing the multiscale spatial organization of fishing vessels on the Patagonia shelf at a given time. These distributions hold an incredible amount of information that we utilize to develop spatial anomaly indexes indicative of IUU activity.

To monitor the spatial organization of fleets over time, as they potentially respond to nearby illegal activities, we imagine a situation in which a real-time analysis is required. As new AIS data is made available, there is a need to compute a spatial anomaly index for that time period. To allow for this, we perform a retrospective analysis wherein for a given leading time period, it's between vessel distance distribution is compared with those from a specific lagged time period. This lagged time period is dependent on two factors: 1) the inherent memory of the system and 2) the typical timescale of IUU activities. The former is important because marine social-ecological systems are inherently non-stationary. As a consequence, an "anomaly" should be relative to a recent time period. The latter is important because if too short a timescale is chosen then a relatively prolonged period of IUU activity will be diagnosed as "normal". For the 2016 and 2018b IUU events, this lagged time period was chosen as 8 days and for the 2018a event, 3 days. These choices were made by iteratively exploring the results of our analysis and choosing time periods that best captured the non-stationarity of the system, as well as the timescales of each IUU event.

Once the lagged timescale was chosen, each between vessel distance distribution was compared with those from the respective lagged time period. This comparison was made using the Kolmogorov-Smirnov (KS) statistic, which measures the distance between two PDFs. The result is a set of (lagged) KS statistics for each time period. To summarize this lagged information, we computed various moments over these sets. In particular, we computed the mean lagged KS statistic per time period, as well as the kurtosis. These each provide information about how anomalous a given time period is based upon the geospatial organization of vessels in the region. 

The last step is to compute the significance of these anomaly indexes. We do so by repeating the calculation of the two anomaly indexes (the mean and kurtosis of the lagged KS statistic sets) but for three null periods. These null periods are those immediately after the 2016 and 2018 IUU events. They were chosen because they are the least likely to have IUU events: the challenge here is to precisely identify times when no illegal activities were conducted, which is difficult because reporting of illegal activities is not accurate, and we could not simply choose time periods at random. As a consequence, we chose the periods after the 2016 and 2018 events because after the preceding naval/coast-guard intervention, any further illegal activity would be highly unlikely during these periods. With these null periods, significant anomalies were then computed as those greater than the 99th percentile of the null distributions.

\section*{RESULTS}

\noindent Between vessel distance distributions change over time (see Fig. 2A for the IUU event that occurred in March 2016, where the  Lu Yan Yuan Yu 010 was caught fishing illegally and ultimately scuttled by the Argentine coast guard, and 2B and C for results for the two other case studies in 2018). Between vessel distance distributions generally have a mode at short spatial distances (<50 km; this is identified by the red bands across time in each panel in Figure 2). This is the spatial scale of local fleets, as can be seen on the map in Fig. 1A, which aggregate around specific fishing hotspots (\cite{sabatini2012}). In addition to this local structure, there is spatial structure in the tails of these between vessel distance distributions, and changes through time can be observed, potentially revealing anomalous spatial behaviors in legally operating vessels (those that transmit AIS data) as they respond to nearby vessels committing illegal fishing (Fig. 2; red lines denote the approximate time of IUU activity). However, from visual inspection of these between vessel distributions through time, it is not immediately clear if there is a spatial signature relating to IUU activity.

To quantitatively and objectively find a signal of IUU activity we compared every between vessel distance distribution with those from a lagged set, using the Kolmogorov-Smirnov statistic. The lagged KS statistics through time, for each IUU case-study, can be seen in Figure 3. In each we see a strong signal of the IUU events: the KS statistic rises around the time of each event (Fig. 3, black vertical lines), identifying that as these events occur, the spatial organization of vessels in the region becomes quite distinct from previous times. Notably, there is nuance to these changes, with the 2016 event producing a pronounced and relatively long period of elevated KS statistic values, relative to the two 2018 events. Interestingly, this rise in the KS statistic in 2016 occurs after the IUU event, whereas there appears to be an elevated signal proceeding the events in 2018, especially 2018a (Fig. 3B).

For every time period, the lagged KS statistic sets were summarized using the mean and kurtosis moments. These define two spatial anomaly indexes (Figures 4 and 5 respectively). The mean anomaly index identifies each IUU event quite clearly (Fig. 4): for each, there are elevated levels of this anomaly index during the period of illegal activity. Interestingly, for the 2016 event the mean anomaly index peaks after the reported date of 2016 IUU event; for the 2018a event, the anomaly peaks approximately at the time of the IUU event, and for the 2018b event, the anomaly peak occurs before the event. This suggests that anomalous spatial behavior can precede the ultimate intervention of the illegal activity.

This is mirrored by the second spatial anomaly index: the kurtosis of the lagged KS statistic sets (Fig. 5). This kurtosis anomaly index shows an even clearer precursor signal for all case-studies. Furthermore, for the 2018b event this index identifies anomalous geospatial patterns in times well before the period IUU activity. This is evidence to suggest that while the other precursor signals might be indicative of the impact of a coast-guard intervention on the geospatial organization of legally operating and visible fleets, these early-warning signals seen in 2018b may be indicative of the response of the "good" vessels to that actual IUU activity committed by dark fleets.

\section*{DISCUSSION}

\noindent To detect IUU fishing quickly, and even make predictions of where it is likely to happen, a new data analytic approach was developed based on concepts from complex systems, specifically multiscale anomaly detection. The approach uses a novel anomaly detection algorithm to identify spatial behaviors exhibited by legally operating and visible (through the transmission of AIS location data) vessels, as they respond to nearby illegal activities. Specifically, changes in between vessel distance distributions were quantified using the Kolmogorov-Smirnov statistic. Then, a moving window was used to emulate a real-time situation in which an anomaly index –moments from the lagged KS statistic sets– is computed as new data arrives. Two anomaly indexes were explored: the mean and kurtosis of the lagged KS statistic sets. These indexes identified anomalous periods, based on the geospatial organization of vessels in the Patagonia shelf region, occurring before, during and after the time of known historical IUU events. Importantly, precursor signals were identified, suggesting the possibility that legally operating vessels do respond anomalously in response to nearby IUU activity. This highlights an opportunity to advance predictive data analytics for IUU activities at sea, globally and in near real-time, and not just retrospective analyses. This has immediate utility marine conservation and fisheries management.

This method for IUU detection is distinct from other approaches, which have to date focused solely on the spatial dynamics of vessels one at a time. In contrast, our approach makes use of the spatial characteristics of whole fleets and assumes that any illegal activity will be revealed by the anomalous spatial behavior of nearby vessels. The utility of this approach was verified for the Patagonia Shelf, a highly productive and busy area for fisheries. We chose this area for a few reasons: first, this area has experienced in the past and continues to suffer chronic exposure to IUU fishing. Second, this is also a congested oceanic area, meaning that at any given time there are numerous vessels both fishing, shipping and recreational operating in the area. This is important because for our spatial anomaly detection algorithm to work, there needs to be vessels proximate to the IUU event. This is, on the one hand, an opportunity, for most of the world's fisheries operate in only a fraction of the world's oceans (albeit a large fraction), and recent work has shown that in most of these fishing regions, there are often many vessels from numerous nations (\cite{kroodsma2018}). However, this is also a major challenge, for there remain many areas of the world's oceans where IUU activity occurs in relatively isolated and remote locations. For example, recent work has highlighted the high seas (i.e. areas far from the exclusive economic zones) as areas where illegal activities occur constantly (\cite{cullis2010}). This means that there exists a key spatial scale at which people/vessels do not react to nearby illegal activities, and at which our spatial anomaly detection algorithm will cease to be useful. This spatial scale is yet unknown, and will also be location and event specific, highlighting further need to examine the spatial and temporal scales of the spatial behavior of vessels. 

New and improved methods for detecting and predicting IUU fishing will help us achieve Sustainable Development Goal 14 (life below water) in one major way; it will improve the accuracy of our estimates of aggregate fishing pressure on fish stocks. Currently, impacts of IUU fishing can be accounted for through indirect means (\cite{pauly2016}), and at coarse spatial scales (i.e. at the scale of exclusive economic zones). Even though limited, results from these assessments suggest that globally IUU fishing accounts for 20\% of total fish catch (\cite{sumaila2006}; \cite{stimson2018}). This information has motivated nations to combat illegal fishing, and the next step is to improve the operational capacity for (maritime) law enforcement groups (i.e. coast guards) through finer-resolution data products. Most current analyses of vessel location data provide information on potential dark-fleet activity after it has occurred. In order to curb illegal fishing, and as a consequence overharvesting of fish stocks, pre-emptive capacity is required. This is what our analysis provides, and although only demonstrated for the Patagonia shelf for a limited number of case studies, vessel location data are now global, and hence, there is an opportunity to monitor anomalous spatial behaviors across the world's oceans. Before doing so, two major caveats to note are 1) we only performed our analysis on three incidences of illegal fishing, this is well below what would normally be necessary to verify the accuracy and precision of an algorithm and 2) our approach does not confirm IUU activity, rather it is suggestive of dark fleet IUU activity, which can be used to direct the attention of coast guards and navies for example.

An important benefit of our approach is that it requires only minimal spatial information, specifically the algorithm only needs the location of vessels through time. Many current approaches to detecting IUU fishing rely on additional information like the vessel Maritime Mobile Service Identity (MMSI) number and the historical trajectory of vessels. While a relatively straightforward goal, in practice it is very difficult tracking vessels from raw AIS data. Spoofing is rife (by our estimates, the raw AIS data that we worked with included 12-20\% of vessels spoofing daily through changing their MMSI number). Methods have been applied to deal with this (e.g. \cite{patroumpas2017}), and some successes have been made, but it remains a challenge. Another route to detecting IUU fishing is, once a dark fleet vessel has been identified (by their MMSI number say), to then trace their spatial history and track them forward in time. The full spatiotemporal trajectory of vessels reveals rich information about their past and present activities, and this has been used to great effect to help in the detection of IUU fishing and can be used to capture dark fleet vessels when they come to port (\cite{miller2016}). However, there are again challenges to associating with tracking vessels through time (primarily because of spoofing). Our approach needs none of this. With simply the location of vessels in a given spatial region at a given time, we can infer anomalous multiscale fleet behavior, which we show can be associated with illegal activities. This approach is agnostic to vessel identity and a vessel's spatial history. As a consequence, the information mined from these data makes the least assumptions about the vessels and regions being studied and is thus arguably more robust.

Monitoring the changing multiscale spatial behavior of fishing fleets has utility beyond detecting and predicting illegal fishing. A variety of bad things happen at sea, including narcotics, arms and human trafficking (\cite{tickler2018}). There is active slavery in many areas of the world's oceans, and geopolitical machinations are implemented through both official (naval) and unofficial actors (e.g. fishing vessels; \cite{fravel2011}). All these kinds of activities happen in and around vessels operating legally by "good" actors. Furthermore, the level of information sharing and contextual knowledge of skippers and crew on vessels is known to be high (\cite{johannes2000}), meaning that they know what is happening around them. Harnessing this knowledge-capital is an opportunity for dealing with maritime illegal activities broadly defined: given this analysis here of illegal fishing, it is likely that other illegal activities committed by "dark vessels" will lead to some form of anomalous spatial behavior in other nearby vessels. Indeed, it would be fascinating to create a typology of anomalous spatial behaviors, with different archetypes associated with different illegal activities. The typology could, for example, include descriptions of whether vessels group together, or spread apart, or create spatial sub-modules, and if so, if there is a typical spatial scale of clustering. This information could inform heuristics for faster identification of IUU activity; i.e. if observed fleets are exhibiting spatial behaviors X and Y, then illegal activity Z could be occurring nearby.

An important and obvious next step in our approach to detecting and predicting IUU activity is to improve upon the algorithm. Between vessel distributions do well at capturing the spatial organization of fleets, however, there are many other ways in which this could have been done. From a mathematical perspective, this simple first step is to put a metric/weight on the edges defining interactions between vessels. Spatial distance is an obvious choice, but other metrics could have been derived from correlations in heading or velocity for example. With different measures of association between vessels, different spatial behavioral anomalies could be detected/classified. Furthermore, methods from the study of complex systems, specifically collective behavior in nature (\cite{rosenthal2015}) and people (\cite{cattuto2010}), have used such measures of association between agents comprising the system, in conjunction with dimensional reduction/denoising algorithms, to identify signals of change otherwise lost in the noise (\cite{jolliffe2011, coifman2004}). These approaches have been applied to big spatial datasets to find a signal of anomalous behaviors, and it is likely a fruitful (and key) next step in the continued development of IUU detection and prediction algorithms.

\section*{CONCLUSION}

\noindent To conclude, combating IUU fishing is key to improving the sustainability of fisheries around the world, and to secure the health of our oceans now and in the future. Important secondary outcomes are diminished geopolitical tension and economic impacts on local economies. Current IUU monitoring relies on detecting spoofing in vessel location records, which is retrospective and confined to identifying anomalous spatial behavior of vessels one at a time. Both are useful but limited in their ability to inform enforcement agencies like coast guards about the occurrence of IUU activity. To overcome these challenges, we have developed an approach to IUU detection, and even prediction, that examines the multiscale spatial behavior of whole fleets, those groups of vessels that are visible and are legally operating, as they respond anomalously to nearby dark fleet activity.  We have shown that this approach identifies consistent precursor signals of IUU activity for a number of known IUU events on the Patagonia shelf, as well as clear identification of the ultimate intervention by the coast-guard. This is a promising new tool for detecting, predicting and deterring IUU fishing. Future research will focus on expanding the analysis to other geographic regions, and for other illegal activities, not just fishing, for example, narcotics trafficking and piracy. If the "good" vessels and fleets continually exhibit anomalous multiscale behaviors across regions and events, then it is possible to create near real-time and global indicators of IUU activity, broadly defined. This new information will improve enforcement of maritime laws, and ultimately the sustainability of our seas.
\newpage
\bibliographystyle{plain}

\newpage
\newpage

\begin{table}
\centering
\caption{Description of Patagonia Shelf IUU Events}
    \begin{tabular}{|C{16mm}|C{16mm}|c|c|}
        \hline
        Date & Fleets \newline involved & Context & Data \\
        \hline 
        March 15, 2016\textsuperscript{1} & China, \newline Argentina & \multicolumn{1}{m{6cm}|}{A Chinese fishing vessel, the Lu Yan Yuan Yu 010, was fishing in Argentina's exclusive economic zone. The Argentina Navy fired warning shots which resulted in the Chinese fishing vessel attempting to crash with the Navy vessel. The Argentine Navy opened fire and sunk the vessel. Four members were rescued by the Navy while the others were rescued by other Chinese fishing vessels. } & \multicolumn{1}{m{6cm}|}{The time scale includes observations from March 1 - March 31 2016 and the spatial scale isolates the ocean region outside of Puerto Madryn along the Argentina coast. The final subset of the data provides 749 unique vessels in 1.263 million square kilometers (1,000.76 km x 1,262.30 km) from March 1 - March 31 2016 for a balanced panel of 2,786,280 observations.}   \\
        \hline
        February 2, 2018\textsuperscript{2} & Spain, \newline Argentina & \multicolumn{1}{m{6cm}|}{A Spanish fishing vessel, the Playa Pesmar Uno, was intercepted by the Argentina Navy for illegally operating in Argentina's EEZ which resulted in a heavy fine.} & \multicolumn{1}{m{6cm}|}{The time scale includes observations from January 15 - February 15 2018 and the spatial scale isolates the ocean region outside of Puerto Madryn along the Argentina coast. The final subset of the data provides 777 unique vessels in 1.263 million square kilometers (1,000.76 km x 1,262.30 km) from January 15 - February 15 2018 for a balanced panel of 2,983,680 observations.}   \\
        \hline
        February 21, 2018\textsuperscript{3} & China, \newline Argentina & \multicolumn{1}{m{6cm}|}{The Argentina Navy warned a Chinese Fishing Vessel, the Jing Yuan 626, to halt because they were operating inside their exclusive economic zone. The vessel ignored the warning and escaped into international waters. Additional Chinese fishing vessels attempted to collide with the Navy vessel to intervene in the chase. The Argentina Navy was unable to capture the Chinese fishing vessels and issued orders for the capture of the five vessels.} & \multicolumn{1}{m{6cm}|}{The time scale includes observations from February 5 - March 10 2018 and the spatial scale isolates the ocean region outside of Puerto Madryn along the Argentina coast. The final subset of the data provides 828 unique vessels in 1.263 million square kilometers (1,000.76 km x 1,262.30 km) from February 5 - March 10 2018 for a balanced panel of 3,378,240 observations.}   \\
        \hline

    \end{tabular}
    \newline
    \parbox{6.5in}{
    \footnotesize
    \textsuperscript{1} \text{www.cnn.com/2016/03/15/americas/argentina-chinese-fishing-vessel/index.html}
    \newline
    \textsuperscript{2} \text{www.laht.com/article.asp?CategoryId=14093\&ArticleId=2450374}
    \newline
    \textsuperscript{3} \text{www.reuters.com/article/us-argentina-china-fishing/argentina-calls-for-capture-of-five-chinese-fishing-boats-idUSKCN1GK35T}
    }
\end{table}
\newpage
\captionsetup{labelfont=bf}

\begin{figure}[]
    \centering
    \includegraphics[height = 6in]{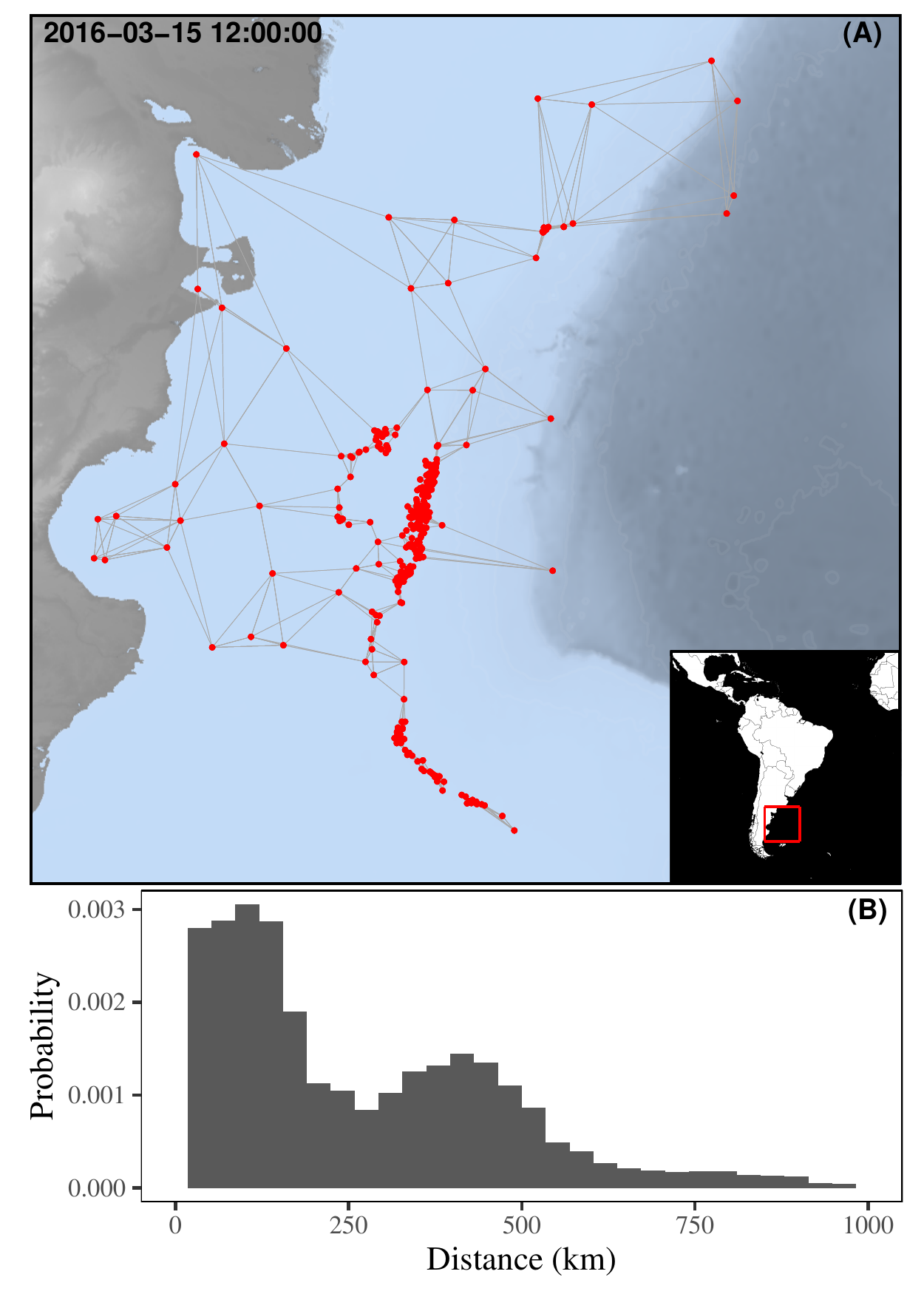}
        \caption[]{\textbf{}{\footnotesize The map shows the location for each vessel on March 15, 2016 at 12PM, with lines connecting nearest five neighbors (chosen for visual clarity). This highlights the spatial network of interactions defining a complex fisheries system. B) A between vessel distance distribution (km) for all vessels. These between vessel distance distributions are used to characterize the multiscale spatial organization of fleets operating in this region. Essentially, changes in between vessel distance distributions reveal IUU events.
}}
    \label{fig:figure1}
    \end{figure}  

    \begin{figure}[]
    \centering
    \includegraphics[height = 6in]{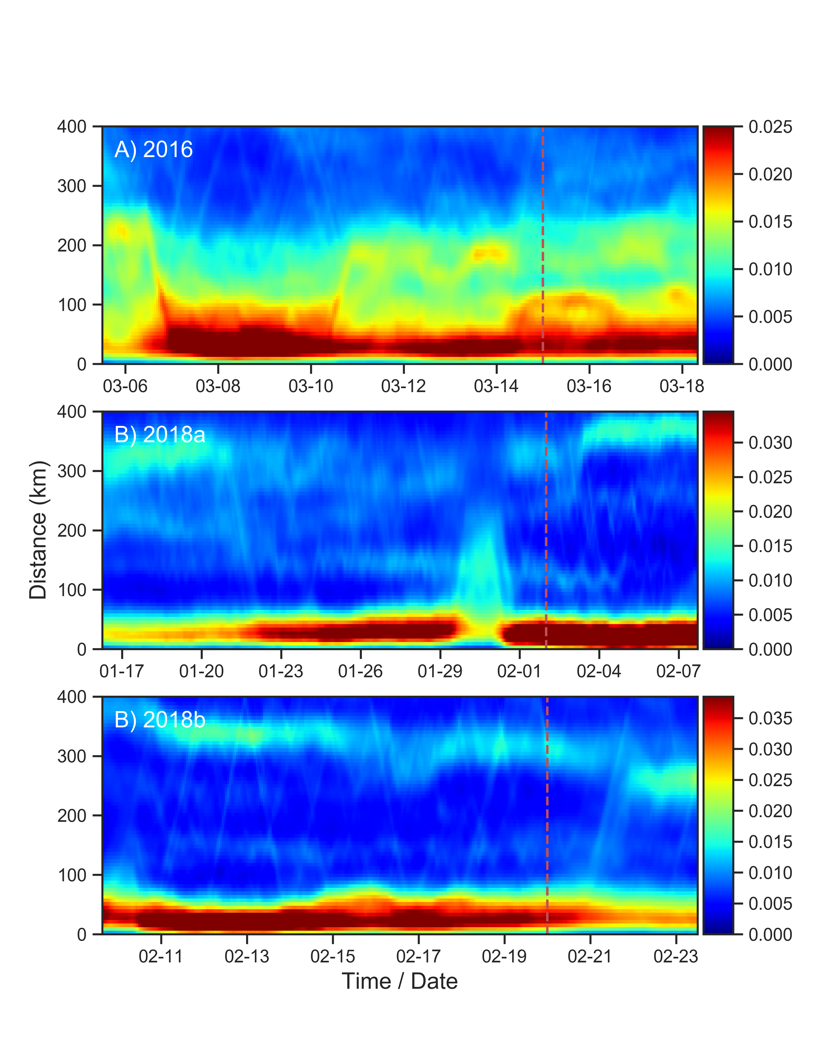}
        \caption[]{\textbf{}{\footnotesize Between vessel distance probability density functions (color scale) through time for three time periods where illegal fishing occurred on the Patagonia shelf: 2016, 2018a and 2018b (see Table 1 for details of these events; red lines in each panel). These between vessel distance distributions have a distinct mode at short distances. This identifies the scale of local fleets in the region. But there exist other minor modes too, identifying smaller fleets operating in the region. Notably, these distributions change shape over time. These changes in the multiscale geospatial organization of vessels in the region can be used to identify anomalous behaviors indicative of illegal activities.
}}
    \label{fig:figure2}
    \end{figure}  

\begin{figure}[]
    \centering
    \includegraphics[height = 6in]{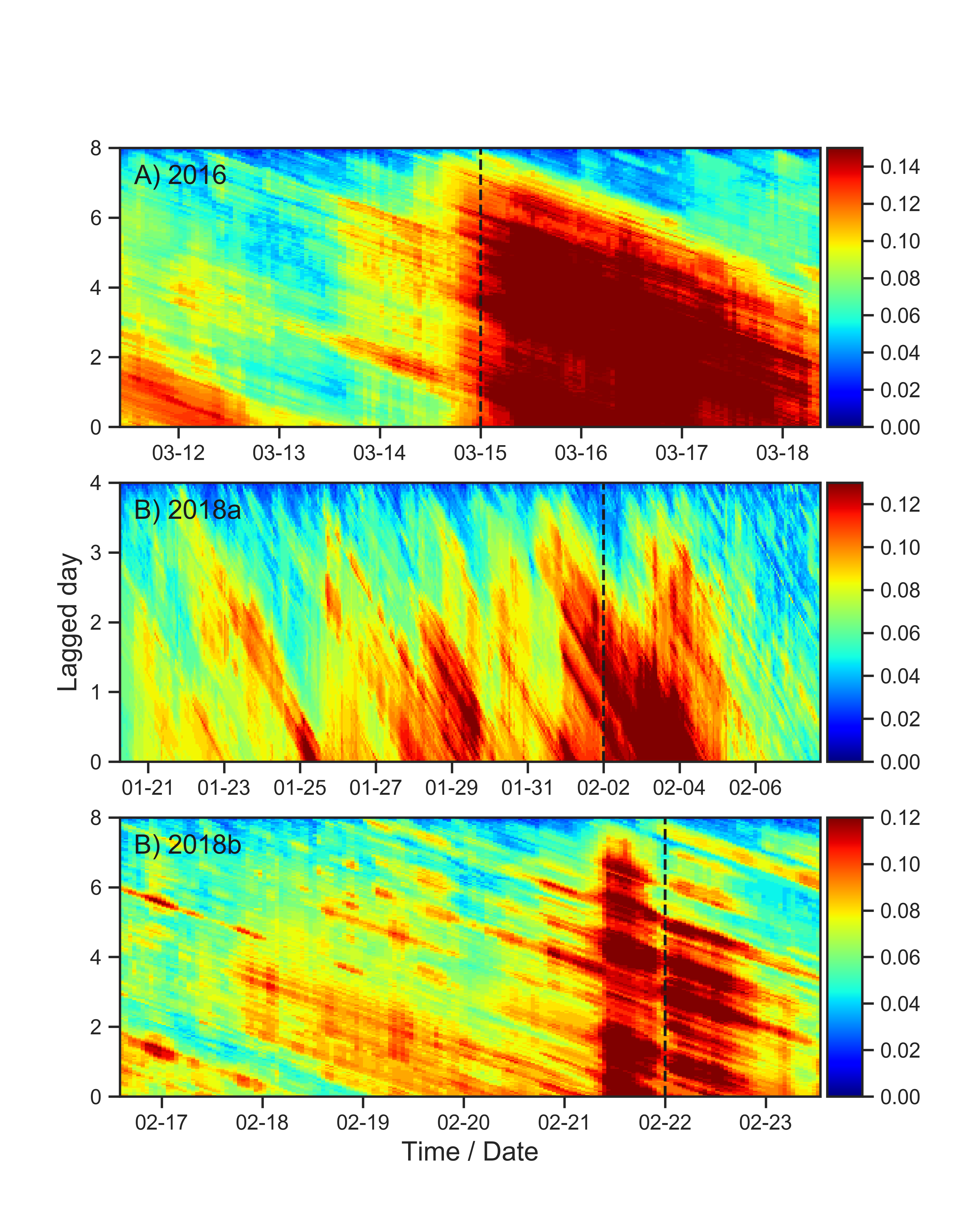}
        \caption[]{\textbf{}{\footnotesize A moving window is used to compare a given time period, based on its between vessel distance distribution (the horizontal axis), with those from a specific lagged period (the vertical axis). The comparison is made using the Kolmogorov-Smirnov (KS) statistic (i.e. the color scale). The KS statistic essentially captures how different a given time period is, based on the geospatial organization of vessels, compared to other lagged times. For each IUU event (black vertical lines), we observe an increase in the lagged KS statistic. This highlights that the geospatial organization of vessels in and around an illegal activity is relatively anomalous. }}
    \label{fig:fig3}
    \end{figure}

\begin{figure}[]
\centering
\includegraphics[height = 7in]{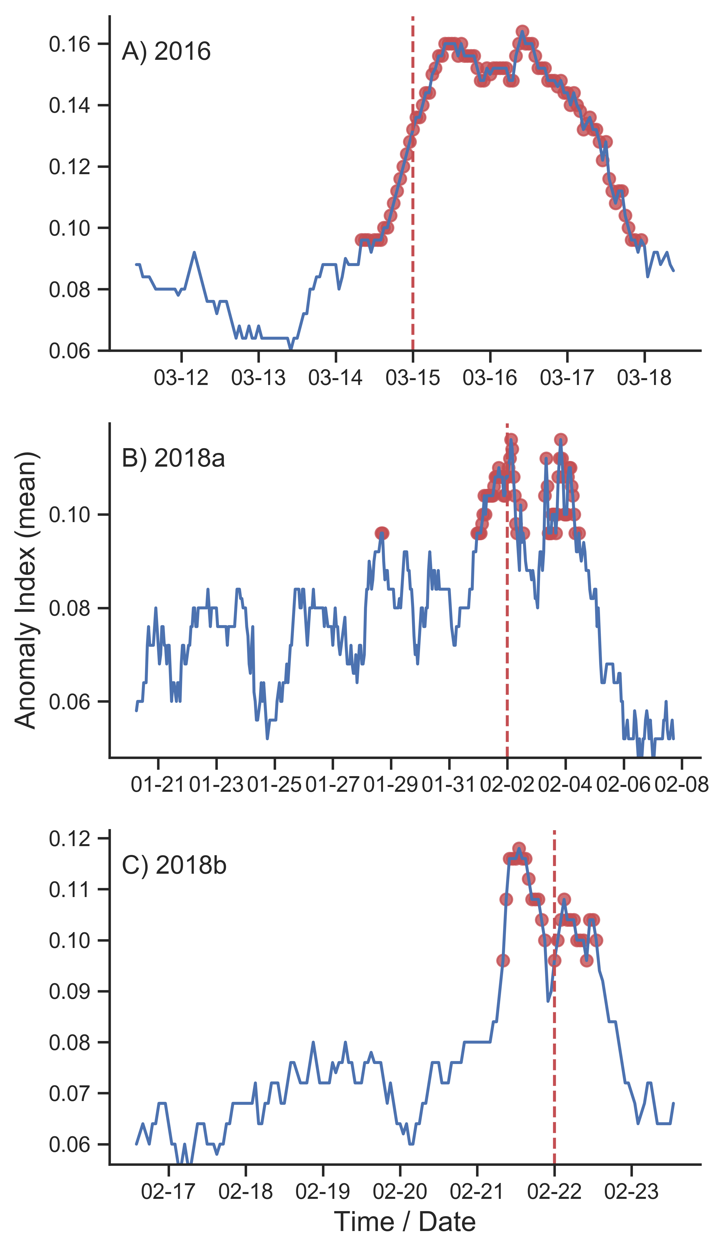}
    \caption[]{\textbf{}{\footnotesize Moments from the lagged KS statistic sets can be used to create anomaly indexes. Here, the mean lagged KS statistic is plotted over time. For each IUU event (vertical red lines) this anomaly index peaks immediately before, during and after the events. "Significant" anomalous times (red markers) are identified by comparing values to those created from null periods where IUU activities were suspected to not have occurred. These significant anomalies again all fall on the times when IUU events occurred, highlighting that this method can be used to identify when these kinds of events occurred. Importantly, precursor signals are observed, suggesting that predicting IUU activities may be possible.}}
\label{fig:figure4}
\end{figure}

\begin{figure}[]
\centering
\includegraphics[height = 7in]{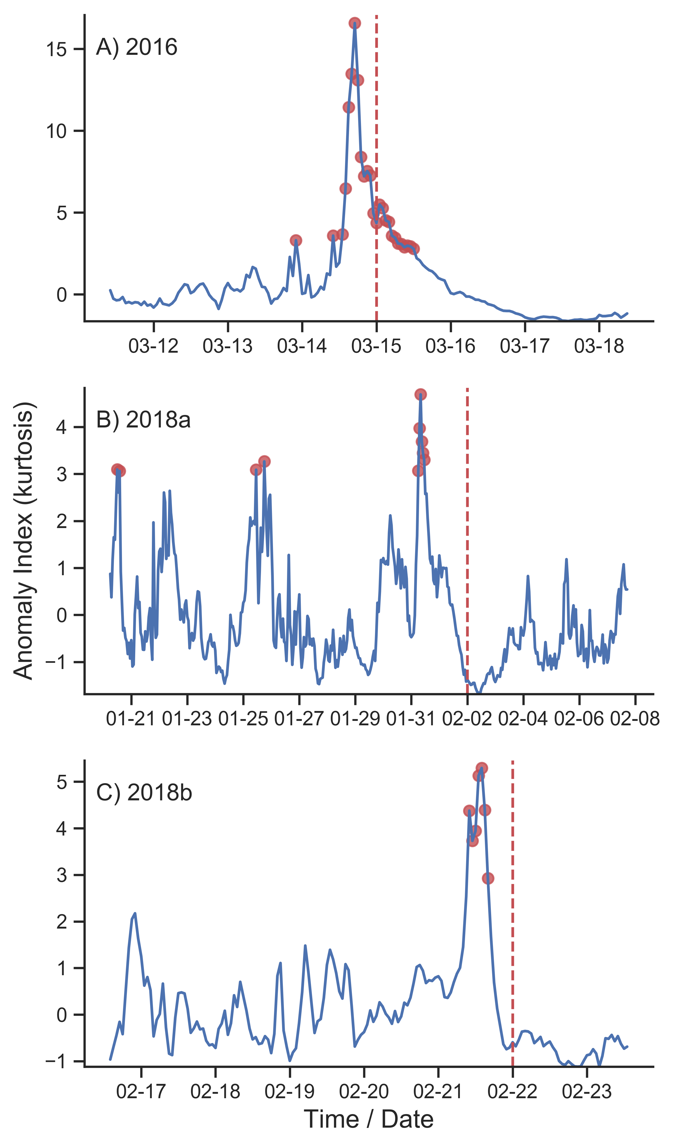}
    \caption[width = 6in]{\textbf{}{\footnotesize Moments from the lagged KS statistic distributions can be used to create anomaly indexes. Here, the kurtosis of the lagged KS statistic sets is plotted over time. "Significant" anomalous times (red markers) are identified by comparing values to those created from null periods where IUU activities were suspected to not have occurred. This anomaly index reveals strong precursor signals before all IUU events. This signal may be indicative of the naval/coast-guard intervention, but potentially more powerful, it may also be indicative of IUU activity itself, especially in (B) the 2018a case-study, where the precursor anomalies occur well before the intervention. 
}}
\label{fig:figure5}
\end{figure}  
\end{document}